\documentclass[submission,copyright,creativecommons]{eptcs}
\providecommand{\event}{sairp} 
\usepackage{breakurl}             

 \newtheorem{thm}{Theorem}[section]

 \newtheorem{prop}[thm]{Proposition}
 \newtheorem{defn}[thm]{Definition}
 \newtheorem{rem}[thm]{Remark}
 \newtheorem{note}[thm]{Notation}
 \newtheorem{exam}[thm]{Example}
 \newtheorem{ques}[thm]{Question}
 \newtheorem{const}[thm]{Construction}
 
 \newtheorem{ans}[thm]{Answer}
 
 \newtheorem{term}[thm]{Terminology}

\title{Formal Contexts, Formal Concept Analysis, and Galois Connections}
\author{Jeffrey T. Denniston
\institute{Department of Mathematical Sciences \\Kent State University \\Kent, Ohio,
USA 44242}
\email{jdennist@kent.edu}
\and
Austin Melton
\institute{Departments of Computer Science and Mathematical Sciences \\
Kent State University \\Kent, Ohio,
USA 44242}
\email{amelton@kent.edu}
\and
Stephen E. Rodabaugh
\institute{College of Science, Technology, Engineering,
Mathematics (STEM) \\Youngstown State University \\Youngstown, OH, USA
44555-3347}
\email{serodabaugh@ysu.edu}
}
\def\titlerunning{Formal Contexts \& Galois Connections}
\def\authorrunning{J.T. Denniston, A. Melton \& S.E. Rodabaugh}
\begin{document}
\maketitle

\begin{abstract}
Formal concept analysis (FCA) is built on a special type of
Galois connections called polarities.  We present new
results in formal concept analysis and in Galois connections
by presenting new Galois connection results and then applying
these to formal concept analysis.  We also approach FCA
from the perspective of collections of formal contexts.
Usually, when doing FCA, a formal context is fixed.  We are
interested in comparing formal contexts and asking what
criteria should be used when determining when one formal
context is better than another formal context.  Interestingly,
we address this issue by studying sets of polarities.
\end{abstract}

\section{Formal Concept Analysis and Order-Reversing Galois Connections}
\label{sect-fca-gc}

We study formal concept analysis (FCA) from a ``larger''
perspective than is commonly done.  We emphasize
formal contexts.  For example,
we are interested in questions such as if we are working with a
given formal context $\mathcal{K}$, that is, we are working with a set of objects
$G$, a set of properties $M$, and a relation $R \subset G \times M$,
what do we do if we want to replace $\mathcal{K}$ with a better
formal context.  Of course, this raises the question: what makes
one formal context better than another formal context.  We address this
question in multiple ways.

\smallskip

We look at sets of formal contexts such that each formal context
has the same set of objects $G$ and the same set of properties $M$,
and we define orderings on these sets of formal contexts.  Each of
these orderings has a mathematical motivation.
We also
look at a category in which the objects are formal contexts.
This category is rich and well structured.
All our results come from using the strong
relationship between formal concept analysis and Galois
connections.  In fact, in getting new ideas and results for
FCA and formal contexts, we also get new results for
Galois connections.

\smallskip

Formal concept analysis (FCA) consists of methods
to analyze data and represent knowledge and is usually done from the perspective
of a single formal context.  The methods and tools used in FCA
are applicable for all formal contexts, but usually one does
not look at FCA from the perspective of a
collection of formal contexts.  In this paper, we look at FCA from
the perspective of ordered collections of formal
contexts and from the
perspective of a category of formal contexts.  Our perspective
also differs from that of many FCA studies because we think of
formal contexts not so much in terms of their usual presentation
but in terms of their effective structure.  The usual presentation
of a formal context is as two sets---a set of objects and a set of
properties which the objects may have---and a relation which relates
or connects an object to the properties which hold for that object.
Though this is the usual presentation, what makes FCA useful in
data analysis and knowledge representation is the Galois
connections each of which is determined by a relation and
is defined between
the powerset of the set of objects and the powerset of the
set of properties.
Each of these Galois connections
clusters sets of objects in the powerset of objects and clusters
sets of attributes in the powerset
of attributes, and the Galois connection also naturally ``connects'' clusters from
the two powersets.
Thus, we organize our study of FCA
in the context or setting of Galois connections.

\smallskip

In this first section, we introduce FCA and Galois connections, and
we review properties of each.  The review emphasizes that the
effective structure of a formal context is the accompanying
Galois connection.  In Section \ref{sect-orderings}, we define
orderings on sets of Galois connections, and these orderings
also order our sets of formal contexts.  In Section \ref{sect-cat},
we present our category of formal contexts and show how this category
embeds into and, in fact, is structured within a larger category of
Galois connections.  Section \ref{sect-example} has an FCA example which
involves the semantic web and which shows the importance of our
goals in studying FCA from a perspective which emphasizes formal contexts.
In our conclusion, Section \ref{sect-con}, we note that we
raise several questions, and we mention
possible areas for related future research.

\begin{defn} \rm
A \emph{formal context} is an ordered triple $(G,M,R)$ where $G$ is a set
of objects, $M$ is a set of attributes, and $R$ is a relation from $G$ to $%
M$, i.e., $R\subset G\times M$.
\end{defn}


\begin{defn} \rm
A \emph{Galois connection} is an ordered quadruple $(f,(P,\leq),
(Q,\sqsubseteq ),g)$ such that $(P,\leq )$ and $(Q,\sqsubseteq )$ are
partially ordered sets, and $f:P\rightarrow Q$ and $g:Q\rightarrow P$ are
order-reversing functions such that for each $p\in P$, $p\leq gf(p)$ and for
each $q\in Q$, $q\sqsubseteq fg(q)$.
\end{defn}

\begin{defn} \rm
\label{defn-alt-gc}
(Alternate Definition) A \emph{Galois connection} is an ordered quadruple $%
(f,(P,\leq ),(Q,\sqsubseteq),g)$ such that $(P,\leq )$ and $(Q,\sqsubseteq)
$ are partially ordered sets, and for each $p\in P$ and $q\in Q$, $p\leq g(q)
$ if and only if $q\sqsubseteq f(p)$.
\end{defn}

Galois connections may be defined with order-reversing or order-preserving
functions. They were originally defined with order-reversing functions
by G. D. Birkhoff \cite{Birkhoff1940, Birkhoff1967} in the
special case in which the partially ordered sets are powersets with the
partial orders being subset inclusion.
Birkhoff called these special order-reversing Galois connections
\emph{polarities}. Subsequently, O. Ore \cite{Ore1944} extended Birkhoff's
notion to arbitrary posets and called them \emph{Galois connexions}. It was J.
Schmidt \cite{Schmidt1953} who retained the name \emph{Galois connections} but
changed the functions to be order-preserving.  A Galois connection with
order-preserving maps is also called an \emph{adjunction}
\cite{Johnstone1982}.

\smallskip

In this paper, we use order-reversing maps, which is standard in FCA.  Thus,
in this paper, a Galois connection is an order-reversing Galois connection.

\begin{note}  \rm
Sometimes for brevity, we may write $(f, g)$ instead of $(f, (P, \leq), (Q,
\sqsubseteq), g)$ for a Galois connection.

We use a superscript $\rightarrow$ on a function symbol to stand for the
corresponding forward powerset function.  For example, if $f: X \rightarrow Y$,
then $f^{\rightarrow}: \wp (X) \rightarrow \wp (Y)$ such that for $A \subset X$,
$f^{\rightarrow}(A) = \{ f(x) : x \in A \} \subset Y$.  This notation
was used by T. S. Blyth in \cite{Blyth1970}.
\end{note}

From the above, it is not clear why one would want to study Galois connections
and FCA in the same paper, and why one would think that new results in one
area should be of interest in the other area.  When Birkhoff defined
the maps in a polarity, he did it in terms of and based on a relation
between the base sets of the two powersets, i.e., he was in an FCA
setting; see Theorem \ref{thm-birk-ops}.  When Ore extended Birkhoff's
polarities to Galois connections, he also showed that there is a
bijection between the set of relations between two sets and the set of
polarities or Galois connections between the powersets of the two
sets.  The bijections are given in Theorem \ref{thm-birk-ops} and in
Construction \ref{const-G}.

The following proposition is well known; see, for example, \cite{EKMS1993}
and \cite{MSS1986}.

\begin{prop} \rm
\label{gc-prop}
Let $(f,(P,\leq ),(Q,\sqsubseteq ),g)$ be a Galois connection.

\begin{enumerate}
\item  $fgf=f$ and $gfg=g$.

\item  The image points are called fixed points. $p\in g^{\rightarrow }(Q)$
if and only if $p=gf(p)$. Likewise, $q\in f^{\rightarrow }(P)$ if and only
if $q=fg(q)$.

\item  $P$ and $Q$ are naturally organized or structured by the fibers of $f$
and $g$, respectively. Each fiber of $f$ contains exactly one point of $%
g^{\rightarrow }(Q)$, and each fiber of $g$ contains exactly one point of $%
f^{\rightarrow }(P)$. The image point in each fiber is the largest element
of the fiber.

\item
\label{prop-leaves-order}
The partition of non-empty fibers of $P$ has the same partially ordered
structure as $g^{\rightarrow }(Q)$, and the partition of non-empty fibers of $Q$ has
the same partially ordered structure as $f^{\rightarrow }(P)$. If $E_1$ and $%
E_2$ are two non-empty fibers or equivalence classes, for example, in $P$, then $%
E_1\leq E_2$ if and only if there exist $p_1\in E_1$ and $p_2\in E_2$ such
that $p_1\leq p_2$.

\item
\label{prop-iso-leaves}
$g^{\rightarrow }(Q)$ and $f^{\rightarrow }(P)$ are anti-isomorphic
partially ordered sets, and $f|_{g^{\rightarrow }(Q)}^{f^{\rightarrow }(P)}:{%
g^{\rightarrow }(Q)}\rightarrow {f^{\rightarrow }(P)}$ and $%
g|_{f^{\rightarrow }(P)}^{g^{\rightarrow }(Q)}:f{^{\rightarrow }(P)}%
\rightarrow g{^{\rightarrow }(Q)}$ are order-reversing bijections. In fact, $%
f|_{g^{\rightarrow }(Q)}^{f^{\rightarrow }(P)}$ and $g|_{f^{\rightarrow
}(P)}^{g^{\rightarrow }(Q)}$ are anti-isomorphic inverses of each other.
Hence, the set of fibers in $P$ and the set of fibers in $Q$ are
anti-isomorphic partially ordered sets.

\item  If $P$ or $Q$ is a [complete] lattice, then so are $g^{\rightarrow
}(Q)$ and $f^{\rightarrow }(P)$. However, $g^{\rightarrow }(Q)$ and $%
f^{\rightarrow }(P)$ need not be sublattices of $P$ and $Q$, respectively.

\item
\label{prop-unique}
$f$ and $g$ uniquely determine each other.
In fact,  for each $p \in P$,
\[f(p) = \bigvee \{q \in Q | p \leq g(q) \},\]
and for each $q \in Q$,
\[g(q) = \bigvee \{p \in P | q \sqsubseteq f(p) \}.\]
\end{enumerate}
\end{prop}

\begin{rem} \rm
In Proposition \ref{gc-prop}, $P$ and $Q$ need not be complete
lattices or even lattices, and thus, joins and meets
in $P$ or $Q$ need not
exist.  However, the joins in item \ref{prop-unique} do exist
when $(f, g)$ is a Galois connection.
\end{rem}

\begin{term} \rm
Based on results in Proposition \ref{gc-prop}, we often use the
following terminology: a fiber of $f$ in $P$ or of $g$ in $Q$ is called a
\emph{leaf}, elements in the same leaf are called \emph{equivalent}, and the
largest element of a leaf is called a \emph{node}. This latter term visually
suggests the fact that a leaf attaches to the subset of fixed points in $P$
or $Q$ by its largest element.  Also, by item \ref{prop-iso-leaves} of Proposition
\ref{gc-prop}, we say that leaf $E$ in $P$ and leaf $F$ in $Q$ are
anti-isomorphic leaves if their nodes are anti-isomorphic nodes.
\end{term}

The following fundamental result from Birkhoff \cite{Birkhoff1940,
Birkhoff1967} is foundational in FCA, in part because
it links the critical information of Proposition \ref{gc-prop} to the
ideas of FCA, as shown throughout this paper.

\begin{thm} \rm
\label{thm-birk-ops}
(Birkhoff Operators). Let $G$ and $M$ be arbitrary sets, and let $R\subset
G\times M$ be a relation. Define $H_{R}:\wp (G)\rightarrow \wp (M)$ and $K_{R}:\wp
(M)\rightarrow \wp (G)$ by
\[
\textrm{for }S\subset G,\,H_{R}(S)=\{m\in M:gRm\,\forall g\in S\}
\]
\[
\textrm{for }T\subset M,\,K_{R}(T)=\{g\in G:gRm\,\forall m\in T\}
\]
$(H_{R},\wp (G),\wp (M),K_{R})$ is a Galois connection where the partial orderings on both $%
\wp (G)$ and $\wp (M)$ are the subset orderings.  When no confusion is likely, we may use
$H$ and $K$ in place of $H_{R}$ and $K_{R}$, respectively.
\end{thm}

As mentioned above, Birkhoff was the first to define a Galois connection, which
he called a polarity.  He defined a polarity in terms of the Birkhoff
operators, given in Theorem \ref{thm-birk-ops}.  Thus, he defined a
Galois connection beginning with two sets and  a relation between them.
It should be noted that for each
Galois connection between powersets,
there exists a relation between the underlying sets
which generates the two order-reversing maps as a
pair of Birkhoff operators.  Thus, for every pair of sets $G$ and $M$, there
is a bijection between the set of Galois connections between
the powersets of $G$ and $M$ and the set of relations from $G$
to $M$; see Ore \cite{Ore1944}.  The bijection from relations to Galois
connections is given in Theorem \ref{thm-birk-ops}, and its inverse
from Galois connections to relations is given in Construction \ref{const-G}.

\smallskip

Most of the following FCA definitions and results may be found in
\cite{GanterWille1996} or \cite{DMR2013}.

\begin{defn}  \rm
\label{fc-defn}
Let $\mathcal{K} :=(G,M,R)$ be a formal context. A \emph{formal concept} of the formal
context is an ordered pair $(A,B)$ with $A\subset G$ and $B\subset M$ such
that $H(A)=B$ and $K(B)=A$. If $(A,B)$ and $(A^{\prime },B^{\prime })$ are
formal concepts of $\mathcal{K}$, then $(A,B)\leq (A^{\prime },B^{\prime })$ if $%
A\subset A^{\prime }$ or, equivalently, if $B^{\prime }\subset B$.
\end{defn}

\begin{defn} \rm
Let $\mathcal{K}:=(G,M,R)$ be a formal context. The set of all formal
concepts of $\mathcal{K}$ with the partial ordering defined in Definition \ref{fc-defn} is
called the \emph{concept lattice of} $\mathcal{K}$.
\end{defn}

\begin{thm} \rm
Let $\mathcal{K}:=(G,M,R)$ be a formal context, and let $(H,\wp (G),\wp
(M),K)$ be the associated Galois connection. The concept lattice of $%
\mathcal{K}$ is a complete lattice; it is isomorphic to $K^{\rightarrow
}(\wp (M))$ and anti-isomorphic to $H^{\rightarrow }(\wp (G))$.
\end{thm}

The following definitions of formal preconcept and formal protoconcept come
from \cite{VormbrockWille2005}.

\begin{defn} \rm
Let $\mathcal{K}:=(G,M,R)$ be a formal context.

\begin{enumerate}
\item  A \emph{formal preconcept} of the formal context is an ordered pair $%
(C,D)$ with $C\subset G$ and $D\subset M$ such that $C\subset K(D)$ or,
equivalently, $D\subset H(C)$.

\item  If $(C,D)$ and $(C^{\prime },D^{\prime })$ are formal preconcepts of
a formal context, then $(C,D)\sqsubseteq (C^{\prime },D^{\prime })$ if $%
C\subset C^{\prime }$ and $D\subset D^{\prime }$. (This partial order on
formal preconcepts is not an extension of the partial order on formal
concepts.)

\item  Let $(C,D)$ be a
formal preconcept of $\mathcal{K}$. The collection of all formal concepts $(A,B)$ such that $%
(C,D)\sqsubseteq (A,B)$ is a subset of the concept lattice of $\mathcal{K}$
and is denoted by $Precon(C,D)$. We order the elements of $Precon(C,D)$ by
the partial ordering on formal concepts, i.e., by $\leq $.
\end{enumerate}
\end{defn}

Formal preconcepts may be thought of as specifying formal concepts in the sense
that formal preconcept $(C,D)$ specifies or ``determines'' formal concept $%
(A,B)$ if $(C,D) \sqsubseteq (A,B)$. However, as $Precon(C,D)$ is a subset
of formal concepts, we use the formal concept partial ordering on $%
Precon(C,D)$.  Also, note that usually a preconcept does not uniquely
specify or determine a formal concept, i.e., $|Precon(C,D)|$ is usually
greater than one, and $Precon(C,D)$ is never empty.  $Precon(C,D)$
always contains at least $(KH(C), H(C))$ and $(K(D), HK(D))$ though
these two formal concepts may be equal.

\begin{prop} \rm
\label{precon-prop}
Let $\mathcal{K}:=(G,M,R)$ be a formal context.

\begin{enumerate}
\item  If $(C,D)$ is a formal preconcept, then $Precon(C,D)$ is itself a
complete lattice with $(KH(C),H(C))$ being the smallest formal concept in $%
Precon(C,D)$ and $(K(D),HK(D))$ being the largest.

\item  For formal preconcept $(C,D),$ every formal concept $(A,B)$ with $%
KH(C)\subset A\subset K(D)$ is in
\linebreak
$Precon(C,D)$.

\item For a formal preconcept $(C,D),$ $|Precon(C,D)| = 1$ if and only if
$(KH(C),H(C)) = (K(D),HK(D))$ if and only if $KH(C) = K(D)$ if and only
if $H(C) = HK(D))$.

\item For a formal preconcept $(C,D),$ $|Precon(C,D)| = 1$ if and only if
$C$ and $D$ are in anti-isomorphic leaves.

\item  If $(C,D)$ be a formal preconcept, then $(C,D)$ is less than or equal
to exactly one formal concept $\left( A,B\right) $ in the preconcept partial
ordering (i.e., there is a unique formal concept $(A,B)$ with $%
(C,D)\sqsubseteq (A,B)$) if and only if $\left( A,B\right) =\left( K\left(
D\right) ,H\left( C\right) \right) .$
\end{enumerate}
\end{prop}

Thinking of formal preconcepts as specifying formal concepts, leads to the
next proposition.

\begin{prop} \rm
Let $\mathcal{K}:=(G,M,R)$ be a formal context, and let $(C,D)$ and $%
(C^{\prime },D^{\prime })$ be formal preconcepts with $(C,D)\sqsubseteq
(C^{\prime },D^{\prime })$. Then $Precon(C^{\prime },D^{\prime })\subset
Precon(C,D)$.  Formal preconcept $(C^{\prime },D^{\prime })$ is maximal
in the $\sqsubseteq$ partial order if and only if $|Precon(C^{\prime },D^{\prime })| = 1$.
\end{prop}

The higher a formal preconcept is in the preconcept partial ordering,
the more specific or precise it is in specifying formal concepts. However,
it is not the case that $(C,D)\sqsubseteq (C^{\prime },D^{\prime })$ if and
only if $Precon(C^{\prime },D^{\prime })\subset Precon(C,D)$.  Moreover, it
may be the case that $(C,D)$ and $(C^{\prime },D^{\prime })$ are both maximal
and $Precon(C,D) = Precon(C^{\prime },D^{\prime })$, but $(C,D)$ and
$(C^{\prime },D^{\prime })$ are not comparable. For example,
if $D=D^{\prime }$ and if $C$ and $C^{\prime }$ are not comparable in the
subset ordering but are in the same leaf, then $%
Precon(C,D)=Precon(C^{\prime },D^{\prime })$ but $(C,D)$ and $(C^{\prime
},D^{\prime })$ are not comparable in the formal preconcept partial order.
Moreover, if the leaf of $C$ and the leaf of $D$ and $D^{\prime }$ are
equivalent, then $(C,D)$ and $(C^{\prime },D^{\prime })$ are both maximal
but still not comparable in the preconcept partial order.

\smallskip

Since the formal preconcepts may be thought of as specifications for the
formal concepts, we propose a pre-order which is defined on the set of
formal preconcepts of a formal context and which more precisely reflects the
thinking that formal preconcepts are specifications for formal concepts.

\begin{defn} \rm
\label{defn-pre-order}
Let $(C,D)$ and $(C^{\prime },D^{\prime })$ be formal
preconcepts of a formal context $\mathcal{K}:=(G,M,R)$. Then $(C,D)\preceq
(C^{\prime },D^{\prime })$ if $Precon(C^{\prime },D^{\prime })\subset
Precon(C,D)$.
\end{defn}

\begin{prop} \rm
\label{prop-new-order}
Let $(C,D)$ and $(C^{\prime },D^{\prime })$ be formal
preconcepts of a formal context $\mathcal{K}:=(G,M,R)$. Then $(C,D)\preceq
(C^{\prime },D^{\prime })$ if and only if $K(D') \subset K(D)$
and $H(C') \subset H(C)$.
\end{prop}

\begin{prop} \rm
Let $\mathcal{K}:=(G,M,R)$ be a formal context, and let $(C,D)$ and $%
(C^{\prime },D^{\prime })$ be formal preconcepts. The following are
equivalent:

\begin{itemize}
\item  $(C,D)$ and $(C^{\prime },D^{\prime })$ are equivalent as formal
preconcepts in the pre-order $\preceq ;$

\item  $(C,D)\preceq (C^{\prime },D^{\prime })$ and $(C^{\prime },D^{\prime
})\preceq (C,D);$

\item  $Precon(C,D)=Precon(C^{\prime },D^{\prime });$

\item  $C$ and $C^{\prime }$ are in the same equivalence class of $\wp (G)$
and $D$ and $D^{\prime }$ are in the same equivalence class of $\wp (M).$
\end{itemize}
\end{prop}

\begin{ques} \rm
What is the partial order gotten from the pre-order
in Definition \ref{defn-pre-order} by
equating ordered pairs which are equivalent in the pre-order?
\end{ques}

To help us answer this question, we use the following notation.

\begin{note} \rm
When $(f, (P, \leq), (Q, \sqsubseteq), g)$ is a Galois connection,,
we ${\cal {L}}(P)$ and ${\cal {L}}(Q)$ to denote the leaves of
$P$ and $Q$, respectively.  Thus, ${\cal {L}}(\wp (G))$ and
${\cal {L}}(\wp (M))$ denote the sets of leaves of the
polarity $(H, {\wp (G)}, {\wp (M)}, K)$.
Further, we use $f^{*}: {\cal {L}}(P) \rightarrow {\cal {L}}(Q)$ for the
map which maps each element in ${\cal {L}}(P)$ to its anti-isomorphic leaf
in ${\cal {L}}(Q)$.  Likewise, we have the maps $g^{*}:{\cal {L}}(Q) \rightarrow {\cal {L}}(P)$,
$H^{*}:{\cal {L}}({\wp (G)}) \rightarrow {\cal {L}}({\wp {M}})$,
and $K^{*}:{\cal {L}}({\wp (M)}) \rightarrow {\cal {L}}({\wp (G)})$.
From items \ref{prop-leaves-order} and \ref{prop-iso-leaves}
of Proposition \ref{gc-prop}, we know that $f^{*}$ and $g^{*}$ are
anti-isomorphic inverses of each other.  Similarly, $H^{*}$ and $K^{*}$ are
anti-isomorphic inverses.
When more than one Galois connection is being discussed, we will use
${\cal {L}}(f,P)$ and ${\cal {L}}(g,Q)$ instead of ${\cal {L}}(P)$
and ${\cal {L}}(Q)$, respectively.
\end{note}

\begin{ans} \rm
\label{ans-order}
The resulting partial order is isomorphic to a partial order on a
subset of $\wp (G) \times \wp (M)$.  If $E$ is a leaf in
$\wp (G)$ and $F$ is a leaf in $\wp (M)$, we want $(E,F)$ to be
in this subset of $\wp (G) \times \wp (M)$ if and only if
$E \leq H^{*}(F)$ and $F \leq K^{*}(E)$, where this $\leq$ is
defined in item \ref{prop-leaves-order} of Proposition \ref{gc-prop}.
We let $\cal{GM}$ denote this subset of $\wp (G) \times \wp (M)$.

Saying $(E,F) \in \cal{GM}$ is equivalent to saying
there exists $C$ in $E$ and there exists $D$ in $F$
with $(C,D)$ a formal preconcept in $\mathcal{K} = (G,M,R)$,
and this is equivalent to saying for every $C$ in $E$ and
every $D$ in $F$, $(C,D)$ is a formal preconcept in $\mathcal{K}$.

For $(E,F), (E^{\prime }, F^{\prime }) \in \cal{GM}$,
$(E,F) \leq (E^{\prime }, F^{\prime })$ if
$K^{*}(F^{\prime }) \leq K^{*}(F)$ and
$H^{*}(E^{\prime }) \leq H^{*}(E)$.
Compare with Proposition \ref{prop-new-order}.
\end{ans}


\begin{defn} \rm
Let $(G,M,R)$ be a formal context. A \emph{formal protoconcept} of the
formal context is a formal preconcept $(C,D)$ such that $Precon(C,D)$
contains exactly one formal concept.
\end{defn}

\begin{thm} \rm
\label{proto-thm}
Let $(G,M,R)$ be a formal context; let $(H,K)$ be the associated Galois
connection of Birkhoff operators; and let $C\subset G,\,D\subset M$. The
following are equivalent:

\begin{enumerate}
\item  $\left( C,D\right) $ is a formal protoconcept.

\item  $Precon(C,D)$ contains exactly one formal concept.

\item  $C$ and $D$ are members of anti-isomorphic leaves.

\item  $\left( K\left( D\right) ,H\left( C\right) \right) $ is a formal
concept of $\left( G,M,R\right) .$

\item  $KH\left( C\right) =K\left( D\right) .$

\item  $HK\left( D\right) =H\left( C\right) .$
\end{enumerate}
\end{thm}

Statement (4) of Theorem \ref{proto-thm} justifies the term ``protoconcept''; and in
practice, (5) and (6) seem the most convenient to apply.

\section{Orderings on Sets of Formal Contexts with the Same Underlying Sets}
\label{sect-orderings}

We are interested in sets of formal contexts which have the
the same set of objects and the same set of properties, and in particular,
we are interested in orderings on these sets of formal contexts.
Studying sets of formal contexts with the same sets of objects and properties is equivalent to
studying sets of polarities in which all polarities have the same ``first''
powerset and all have the same ``second'' powerset.  Further, if we think in terms of
Galois connections instead of just polarities, then we can study sets
of Galois connections for which the ``first'' partially ordered
set is, for example, $(P,\leq)$ and the ``second'' partially ordered set is,
for example, $(Q,\sqsubseteq)$.



\subsection{A Partial Ordering on ${\cal G}$}

\begin{note} \rm
Let $(P, \leq)$ and $(Q, \sqsubseteq)$ be partially ordered sets.  We
let ${\cal G}(P,Q)$ or simply ${\cal G}$ denote the set of all Galois connections between
$(P, \leq)$ and $(Q, \sqsubseteq)$.
\end{note}

For our first ordering on ${\cal G}$, we use the partial order on
$Q$ and extend it pointwise to the first maps in the elements of
${\cal G}$.

\begin{defn} \rm
Let $(f_{1}, g_{1}), (f_{2}, g_{2}) \in {\cal G}$.
$(f_{1}, g_{1}) \leq (f_{2}, g_{2})$ if for each $p \in P$,
$f_{1}(p) \sqsubseteq f_{2}(p)$.
\end{defn}

\begin{prop} \rm
\label{prop-g}
Let $(f_{1}, g_{1}), (f_{2}, g_{2}) \in {\cal G}$.
$(f_{1}, g_{1}) \leq (f_{2}, g_{2})$ if and only if
for each $q \in Q$, $g_{1}(q) \leq g_{2}(q)$.
\end{prop}

\noindent
{\bf Proof:} \, Suppose $(f_{1}, g_{1}) \leq (f_{2}, g_{2})$.
Let $q \in Q$.
$g_{1}(q) = \bigvee \{p \in P | q \sqsubseteq f_{1}(p) \}$ and
$g_{2}(q) = \bigvee \{p \in P | q \sqsubseteq f_{2}(p) \}$.  Since
for each $p \in P$, $f_{1}(p) \sqsubseteq f_{2}(p)$, then
$\{p \in P | q \sqsubseteq f_{1}(p) \} \subset
\{p \in P | q  \sqsubseteq f_{2}(p) \}$.  Hence,
for each $q \in Q$,
$\bigvee \{p \in P | q \sqsubseteq f_{1}(p) \} \leq \bigvee \{p \in P | q \sqsubseteq f_{2}(p) \}$,
and $g_{1}(q) \leq g_{2}(q)$.

\smallskip

We proved if for each $p \in P$, $f_{1}(p) \sqsubseteq f_{2}(p)$,
then for each $q \in Q$, $g_{1}(q) \leq g_{2}(q)$.  Since the definition of
Galois connections is symmetric with respect to $f$ and $g$,
then if we have for each $q \in Q$ that $g_{1}(q) \leq g_{2}(q)$, it will follow
for each $p \in P$ that $f_{1}(p) \sqsubseteq f_{2}(p)$, i.e., it follows that
$(f_{1}, g_{1}) \leq (f_{2}, g_{2})$.
$\bullet$


\begin{note} \rm
We use ${\cal P}(G,M)$ to denote ${\cal G}(\wp (G), \wp (M))$.
\end{note}

The ordering $\leq$ on ${\cal G}$ is natural, and since $\sqsubseteq$ is a
partial order on $Q$, then $({\cal G}, \leq)$ is also a partially ordered set.
Interestingly, there is an equally natural way to define $\leq$ on ${\cal P}(G,M)$.
This new definition, see below, emphasizes the standard presentation of formal contexts.

\smallskip

As mentioned above, there is a bijection from the set of all
relations from $G$ to $M$ to the set of all polarities ``built
from'' $G$ and $M$, i.e., to the set of all Galois connections
from $\wp (G)$ to $\wp (M)$.

\smallskip

Let ${\cal R}(G,M)$ be the set of all relations from $G$ to
$M$.  Define ${\cal F} : {\cal R}(G,M) \rightarrow {\cal P}(G,M)$
by ${\cal F} (R) = (H_{R},\wp (G),\wp (M),K_{R})$ as defined in Theorem
\ref{thm-birk-ops}.

\begin{defn} \rm
Define the ordering ${\leq}_{\cal P}$ on ${\cal P}(G,M)$ such that if
$(H_{1}, K_{1}), (H_{2}, K_{2}) \in {\cal P}(G,M)$, then
$(H_{1}, K_{1}) {\leq}_{\cal P} (H_{2}, K_{2})$ if and only if
${\cal F}^{-1}(H_{1}, K_{1}) \subset {\cal F}^{-1}(H_{2}, K_{2})$.
Thus, if $R_{1}$ is the relation from $G$ to $M$ such that
${\cal F} (R_{1}) = (H_{1}, K_{1})$ and $R_{2}$ is the relation
$G$ to $M$ such that
${\cal F} (R_{2}) = (H_{2}, K_{2})$ and $R_{1} \subset R_{2}$,
then $(H_{1}, K_{1}) {\leq}_{\cal P} (H_{2}, K_{2})$.
\end{defn}

\begin{prop} \rm
The partial order $\leq$ on ${\cal G}(\wp (G), \wp (M))$ and the
order ${\leq}_{\cal P}$ on ${\cal P}(G,M)$ are the same.
\end{prop}

\noindent
{\bf Proof:} \, Suppose that $(H_{1}, K_{1}), (H_{2}, K_{2}) \in {\cal P}(G,M)$
with $R_{1} = {\cal F}^{-1}(H_{1}, K_{1})$, and $R_{2} = {\cal F}^{-1}(H_{2}, K_{2})$.
Further suppose that $(H_{1}, K_{1}) {\leq}_{\cal P} (H_{2}, K_{2})$, and
let $S \in {\wp (G)}$.
\[ H_{1}(S)=\{m\in M:g R_{1} m\,\forall g\in S\} \] and
\[ H_{2}(S)=\{m\in M:g R_{2} m\,\forall g\in S\}. \]
Since $R_{1} \subset R_{2}$, then
$\{m\in M:g R_{1} m\,\forall g\in S\} \subset \{m\in M:g R_{2} m\,\forall g\in S\}$.
Thus, $H_{1} \leq H_{2}$, and
$(H_{1}, K_{1}) \leq (H_{2}, K_{2})$.

\smallskip

If we begin with $(H_{1}, K_{1}) \leq (H_{2}, K_{2})$, then for
each $S \in {\wp (G)}$, we will have
$H_{1}(S) \subset H_{2}(S)$.  Thus, in particular, for each
$g \in G$, we have $H_{1}(\{g\}) \subset H_{2}(\{g\})$, and this
implies that
${\cal F}^{-1}(H_{1}, K_{1}) \subset {\cal F}^{-1}(H_{2}, K_{2})$.
(See Construction \ref{const-G}.)
$\bullet$

\smallskip

Since $\leq$ and ${\leq}_{\cal P}$ are the same on ${\cal P}(G,M)$,
we will use $\leq$ for both.

\begin{prop} \rm
Let $(f, (P, \leq), (Q, \sqsubseteq), g) \in {\cal G}$.  If
$(P, \leq)$ and $(Q, \sqsubseteq)$ both have greatest elements,
${\top}_{P}$ and ${\top}_{Q}$, respectively, then $({\cal G}, \leq)$ has
a greatest element.
If $(P, \leq)$ and $(Q, \sqsubseteq)$ both also have least elements,
${\perp}_{P}$ and ${\perp}_{Q}$, respectively, then $({\cal G}, \leq)$ also has
a least element.
\end{prop}

\noindent
{\bf Proof:} \, If $(P, \leq)$ and $(Q, \sqsubseteq)$ both have greatest elements,
then $(f_{\top}, g_{\top})$ defined by
\[ \forall p \in P, \; f_{\top}(p) = {\top}_{Q} \]
and
\[ \forall q \in Q, \; g_{\top}(q) = {\top}_{P} \]
is the greatest element in ${\cal G}$.

\smallskip

If $(P, \leq)$ and $(Q, \sqsubseteq)$ both also have least elements,
then $(f_{\perp}, g_{\perp})$ defined by
\[f_{\perp}(p) = \left\{ \begin{array}{ll}
                       {\perp}_{Q} & \mbox{if $p \not= {\perp}_{P}$} \\
                       {\top}_{Q} & \mbox{if $p = {\perp}_{P}$}
                     \end{array}
             \right. \]
and
\[g_{\perp}(q) = \left\{ \begin{array}{ll}
                       {\perp}_{P} & \mbox{if $q \neq {\perp}_{Q}$} \\
                       {\top}_{P} & \mbox{if $q = {\perp}_{Q}$}
                     \end{array}
             \right. \]
is the least element in ${\cal G}$.
$\bullet$













\begin{const} \rm
\label{const-G}
Before we define another ordering on ${\cal G}$, we want to give the explicit
definition of ${\cal F}^{-1} : {\cal P}(G,M) \rightarrow {\cal R}(G,M)$.
Let $(H, \wp (G), \wp (M), K) \in {\cal P}(G,M)$.
${\cal F}^{-1}(H, K) = (G, M, R)$ where $(g, m) \in R$ if and only if
$ m \in H(\{g\})$ if and only if $g \in K(\{m\})$.
\end{const}



\subsection{Other Orderings on ${\cal G}$}

If $(f, (P, \leq), (Q, \sqsubseteq), g)$ is a Galois connection,
then for each $p \in P$, $p \leq gf(p)$, and for each $q \in Q$,
$q \sqsubseteq fg(q)$.  These inequalities can be given by
$1_{P} \leq gf$ and $1_{Q} \sqsubseteq fg$.  These
inequalities lead one to think of trying to
minimize the differences between $1_{P}$ and $gf$
and between $1_{Q}$ and $fg$, which in turn
leads, among other things, to the following
ordering on $\cal G$.\footnote{This ordering was
suggested by C. Rohwer during a conversation at the University
of Stellenbosch, South Africa.}

\begin{defn} \rm
Let $(f_{1},g_{1}), (f_{2},g_{2}) \in {\cal G}(P,Q)$.  Define
$(f_{1},g_{1}) \preceq (f_{2},g_{2})$ if and only if for each
$p \in P$, $g_{2}f_{2}(p) \leq g_{1}f_{1}(p)$.
\end{defn}

\begin{note} \rm
Let $(f, (P, \leq), (Q, \sqsubseteq), g)$ be a Galois connection.
We use ${\cal {N}}(f,P)$ and ${\cal {N}}(Q,g)$ to denote the sets of
nodes of $P$ and $Q$, i.e., the sets of image points of $g$ and $f$, respectively.
\end{note}

\begin{prop} \rm
\label{prop-ref-part}
Let $(f_{1},g_{1}), (f_{2},g_{2}) \in {\cal G}(P,Q)$.  The following
are equivalent.

\begin{enumerate}

\item
$(f_{1},g_{1}) \preceq (f_{2},g_{2})$

\item
${\cal {N}}(f_{1},P) \subset {\cal {N}}(f_{2},P)$

\item
The partition of $P$ by $f_{2}$ is a refinement of the
partition of $P$ by $f_{1}$, i.e., for each $E_{2} \in {\cal {L}}(f_{2},P)$,
there is an $E_{1} \in {\cal {L}}(f_{1},P)$ such that $E_{2} \subset E_{1}$.

\end{enumerate}
\end{prop}

\noindent
{\bf Proof:} \, $(1. \Rightarrow 2.)$
Let $p^{*} \in {\cal {N}}(f_{1},P)$.  Then
$p^{*} = g_{1}f_{1}(p^{*})$.  Since $p^{*} \leq g_{2}f_{2}(p^{*}) \leq g_{1}f_{1}(p^{*})$,
then $p^{*} = g_{2}f_{2}(p^{*})$, and thus, $p^{*} \in {\cal {N}}(f_{2},P)$.

\smallskip

$(2. \Rightarrow 1.)$
Let $p \in P$.  Then $p \leq g_{1}f_{1}(p)$.  Since both
$f_{2}$ and $g_{2}$ are order-reversing, then
$g_{2} f_{2}$ is order-preserving.  Therefore,
$g_{2} f_{2} (p) \leq g_{2} f_{2}g_{1}f_{1}(p)$.
However, $g_{1}f_{1}(p) \in {\cal {N}}(f_{1},P) \subset {\cal {N}}(f_{2},P)$.
It follows that $g_{2} f_{2}g_{1}f_{1}(p) = g_{1}f_{1}(p)$, and
therefore,  $g_{2} f_{2} (p) \leq g_{1}f_{1}(p)$.

\smallskip

$(1. \Rightarrow 3.)$
To show that ${\cal {L}}(f_{2},P)$ is a refinement of ${\cal {L}}(f_{1},P)$,
we begin with $p_{1}, p_{2} \in P$ with $f_{2}(p_{1}) = f_{2}(p_{2})$, and we
show that $f_{1}(p_{1}) = f_{1}(p_{2})$.  Since $f_{2}(p_{1}) = f_{2}(p_{2})$,
then $g_{2}f_{2}(p_{1}) = g_{2}f_{2}(p_{2})$.  It follows that
$p_{1} \leq g_{2}f_{2}(p_{1}) \leq g_{1}f_{1}(p_{1})$, and therefore,
$f_{1}g_{1}f_{1}(p_{1}) \sqsubseteq f_{1}g_{2}f_{2}(p_{1}) \sqsubseteq f_{1}(p_{1})$.
However, since $f_{1}(p_{1}) = f_{1}g_{1}f_{1}(p_{1})$,
then $f_{1}(p_{1}) = f_{1}g_{2}f_{2}(p_{1})$.  Likewise,
$f_{1}(p_{2}) = f_{1}g_{2}f_{1}(p_{2})$.  Since
$g_{2}f_{2}(p_{1}) = g_{2}f_{2}(p_{2})$, then
$f_{1}g_{2}f_{2}(p_{1}) = f_{1}g_{2}f_{2}(p_{2})$,
and $f_{1}(p_{1}) = f_{1}(p_{2})$.
$\bullet$

\smallskip

$(3. \Rightarrow 1.)$
Let $p \in P$.  Assume $p \in E_{2} \in {\cal {L}}(f_{2},P)$,
and assume $p \in E_{1} \in {\cal {L}}(f_{1},P)$.  Since
$p$ is in both $E_{2}$ and $E_{1}$ and since ${\cal {L}}(f_{2},P)$
refines ${\cal {L}}(f_{1},P)$, then $E_{2} \subset E_{1}$.  Since
$g_{2}f_{2}(p)$ is the largest element in $E_{2}$ and
$g_{1}f_{1}(p)$ is the largest element in $E_{1}$,
then $g_{2}f_{2}(p) \leq g_{1}f_{1}(p)$.
$\bullet$

\smallskip

If $(P, \leq)$ and $(Q, \sqsubseteq)$ have greatest elements,
${\top}_{P}$ and ${\top}_{Q}$, respectively, then the
least element in $({\cal G}, \preceq)$ is $(f,g)$ such that
both $f$ and $g$ map everything to ${\top}_{Q}$ and
${\top}_{P}$, respectively.  Interestingly, this $(f,g)$
is the greatest element in $({\cal G}, \leq)$.

\smallskip

$(f',g')$ is a maximal element in $({\cal G}, \preceq)$ if
$1_{P} = g' f'$.  $({\cal G}, \preceq)$ may have multiple
maximal elements, and if $(f',g')$ and $(f'',g'')$ are both
maximal elements, then $(f',g') \preceq (f'',g'')$ and
$(f'',g'') \preceq (f',g')$.  Thus, $\preceq$ is only a
pre-ordering on $\cal G$.  Example \ref{exam-max} shows that
maximal elements are not unique.

\smallskip

Given the symmetric nature of Galois connections, one might think
that if $(f_{2},g_{2}$ produces a finer partition on $P$ than
does $(f_{1},g_{1})$, then $(f_{2},g_{2})$ would also produce
a finer partition on $Q$.  In other words, one might think that
the statement for every $p \in P$, $g_{2}f_{2}(p) \leq g_{1}f_{1}(p)$
is equivalent to the state for every $q \in Q$,
$f_{2}g_{2}(q) \sqsubseteq f_{1}g_{1}(q)$.  However, as the next
example, Example \ref{exam-no}, shows this is not the case.

\begin{exam} \rm
\label{exam-no}
Let $P = \{1, 2, 3 \}$, and let $Q = \{1, 2, 3, 4\}$ with
the usual ordering on natural numbers.  For $f_{1}$, define
$f_{1}(3) = f_{1}(2) = 2$ and $f_{1}(1) = 4$.  Thus,
$g_{1}$ must be $g_{1}(4) = g_{1}(3) = 1$, and
$g_{1}(2) = g_{1}(1) = 3$.  For $f_{2}$, define
$f_{2}(3) = 1$; $f_{2}(2) = 3$; and $f_{2}(1) = 4$.
Then $g_{2}$ must be $g_{2}(4) = 1$; $g_{2}(3) = g_{2}(2) = 2$;
and $g_{2}(1) = 3$.  We have $(f_{1},g_{1}) \preceq (f_{2},g_{2})$,
i.e., for each $p \in P$, $g_{2}f_{2}(p) \leq g_{1}f_{1}(p)$
and ${\cal L}(f_{2},P)$ refines ${\cal L}(f_{1},P)$, but
$f_{2}g_{2}(2) = 3 \not\leq 2 = f_{1}g_{1}(2)$ and
${\cal L}(g_{2},Q)$ does not refine ${\cal L}(g_{1},Q)$.
\end{exam}

Given that $\preceq$ ``works nicely'' on partitions on $P$
but seemingly unpredictably on partitions on $Q$, we could
rename $\preceq$ to ${\preceq}_{P}$ and define another
pre-order ${\preceq}_{Q}$ which we would define for
$(f_{1}, g_{1}), (f_{2}, g_{2}) \in {\cal G}$ by
$(f_{1}, g_{1}) {\preceq}_{Q} (f_{2}, g_{2})$ if
for each $q \in Q$, $f_{2}g_{2}(q) \sqsubseteq f_{1}g_{1}(q)$.

Further, we could define ${\preceq}_{PQ}$ by
$(f_{1}, g_{1}) {\preceq}_{PQ} (f_{2},g_{2})$
if and only if $(f_{1},g_{1}) {\preceq}_{P} (f_{2},g_{2})$
and
\linebreak
$(f_{1},g_{1}) {\preceq}_{Q} (f_{2},g_{2})$.
It would seem that ${\preceq}_{PQ}$ might have promise
for FCA use because it refines partitions in both
$\wp (G)$ and $\wp (M)$.

The next definition comes from Ore \cite{Ore1944}.

\begin{defn} \rm
Let $(f,P,Q,g)$ be a Galois connection.  $(f,g)$ is said to be
\emph{perfect} if each element in $P$ is a node and each element in $Q$
is a node.
\end{defn}

\begin{prop} \rm
If $(f,g)$ is a perfect Galois connection, then
$(f,g)$ is a maximal element in $({\cal G}, {\preceq}_{PQ})$.
\end{prop}

In the introduction, we mention that studying FCA can help
generate new Galois connection results.  FCA motivated
the defining of ${\preceq}_{P}$
and then of ${\preceq}_{Q}$ and ${\preceq}_{PQ}$.

\smallskip

Given the symmetric of Galois connections in general
and the symmetric of ${\preceq}_{PQ}$, one might think
that ${\preceq}_{PQ}$ would be a partial order.  However,
as Example \ref{exam-max} shows, this is not the case.

\begin{exam} \rm
\label{exam-max}
Let $P$ be the four element set $\{ \perp, a, b, \top \}$,
and define the partial order $\leq$ on $P$ such that
$\perp$ is less than or equal to everything, everything is
less than or equal to $\top$, and $a$ and $b$ are not related.
Let $(Q, \leq) = (P, \leq)$, and define $(f,g)$ and $(f',g')$ such
that $f = g$ where $f(\perp) = \top$, $f(\top) = \perp$,
$f(a) = a$, and $f(b) = b$.

\smallskip

Let $f' = f$ except $f'(a) = b$
and $f'(b) = a$, and let $g' = f'$.  Then $(f,g) \neq (f',g')$,
but $(f,g) {\preceq}_{PQ} (f',g')$ and $(f',g') {\preceq}_{PQ} (f,g)$.  Also, both
$(f,g)$ and $(f',g')$ are maximal elements in $({\cal G}, {\preceq}_{PQ})$.
It is also true that $(P, \leq)$ is isomorphic to the powerset of a two
element set.
\end{exam}


\subsection{Yet Another Ordering on ${\cal G}$}

For this ordering, we need and use results from the next section.
In particular, we
use the category $\bf Gal$ and the fact that $\bf Gal$ is concrete
over ${\bf Set} \times {\bf Set}$ with the forgetful functor
\[ U_{G}: {\bf Gal} \rightarrow {\bf Set} \times {\bf Set} \]
defined by
\[ U_{G}(f, (P, \leq), (Q, \sqsubseteq), g) = (P, Q), \]
where $P$ and $Q$ in the image ordered pair are sets without partial orderings.

\smallskip

From \cite{AHS}, we can define a pre-order on the fibers of $U_{G}$
such that if
$(f_{1}, (P_{1}, {\leq}_{1}), (Q_{1}, {\sqsubseteq}_{1}), g_{1})$ and
$(f_{2}, (P_{2}, {\leq}_{2}), (Q_{2}, {\sqsubseteq}_{2}), g_{2})$
are $\bf Gal$-objects with $U_{G}(f_{1}, g_{1}) = U_{G}(f_{2}, g_{2})$,
i.e., such that $P_{1} = P_{2}$ and $Q_{1} = Q_{2}$, then
$(f_{1}, g_{1}) \sqsubseteq (f_{2}, g_{2})$ if and only if
the ${\bf Set} \times {\bf Set}$ identity
$(1_{P}, 1_{Q}): P \times P \rightarrow Q \times Q$
where $P = P_{1} = P_{2}$ and $Q = Q_{1} = Q_{2}$ is a $\bf Gal$-morphism
\[ (1_{P}, 1_{Q}): (f_{1}, (P, {\leq}_{1}), (Q, {\sqsubseteq}_{1}), g_{1})
\rightarrow (f_{2}, (P, {\leq}_{2}), (Q, {\sqsubseteq}_{2}), g_{2}). \]

\smallskip

Using item \ref{prop-iso-leaves} of Proposition \ref{gc-prop}, we know that
\[ (f_{1}, (P, {\leq}_{1}), (Q, {\sqsubseteq}_{1}), g_{1})
\sqsubseteq (f_{2}, (P, {\leq}_{2}), (Q, {\sqsubseteq}_{2}), g_{2}) \]
if and only if the set of nodes in $(P, {\leq}_{1})$ is a subset of the set of nodes of
$(P, {\leq}_{2})$ and the set of nodes of $(Q, {\sqsubseteq}_{1})$ is a subset of the set of nodes of
$(Q, {\sqsubseteq}_{2})$ and $(1_{P}, 1_{Q})$ is a $\bf Gal$-morphism.

\smallskip

We have defined $\sqsubseteq$ in a categorical setting,
and in fact, a fiber in $U_{G}$ is more complex than $\cal G$.
In $\cal G$, we are assuming that $(P, \leq)$ and $(Q, \sqsubseteq)$
are fixed.  However, for a fiber of $U_{G}$, we only have $P$ and $Q$
fixed; the partial orders are not fixed.  We let $\cal H$ be a fiber of
$U_{G}$.

\begin{rem} \rm
$({\cal H}, \sqsubseteq)$ is an interesting pre-ordered set.  We can define
partial orders $\leq$ and $\sqsubseteq$ on $P$ and $Q$, respectively,
so that both have largest elements.  Then we can define a Galois connection,
$(f, \left(P, \leq \right), (Q, \sqsubseteq), g)$ such that $f$ and $g$ are constant functions
where $f$ maps everything in $P$ to the largest element in $(Q, \sqsubseteq)$
and $g$ maps everything in $Q$ to the largest element in $(P, \leq)$.
$(f, (P, \leq), (Q, \sqsubseteq), g)$ is a minimal element in
$({\cal H}, \sqsubseteq)$ though it may not be a least element.

\smallskip

If $|P| = |Q|$, then we can define anti-isomorphic partial orders
$\leq$ and $\sqsubseteq$ on $P$ and $Q$, respectively.  If $f$
is an anti-isomorphism and if $g = f^{-1}$, then
$(f, (P, \leq), (Q, \sqsubseteq), g)$ is a maximal element in
$({\cal H}, \sqsubseteq)$ though it may not be a greatest element.
\end{rem}

We use what we have learned from this ordering on ${\cal H}$ to define
an ordering on ${\cal G}$.


\begin{defn} \rm
Let $(P, \leq)$ and $(Q, \sqsubseteq)$ be partially ordered sets.
Define $\sqsubseteq$ on ${\cal G}$ by
\[ (f_{1}, (P, \leq), (Q, \sqsubseteq), g_{1})
\sqsubseteq (f_{2}, (P, \leq), (Q, \sqsubseteq), g_{2}) \]
if and only if
\[ {\cal N}(f_{1},P) \subset {\cal N}(f_{2},P) \textrm{ and }
{\cal N}(Q,g_{1}) \subset {\cal N}(Q,g_{2}). \]
\end{defn}

\begin{prop} \rm
The ordered sets $(\cal G, {\preceq}_{PQ})$ and $(\cal G, \sqsubseteq)$ are the
same.
\end{prop}

\noindent
{\bf Proof:} \,  Proposition \ref{prop-ref-part}. $\bullet$


\section{Categories of Formal Contexts and Categories of Galois Connections}
\label{sect-cat}

We define and study two categories of formal contexts.  These categories are
interesting, in part, because, though
the objects of these categories are formal contexts, the morphisms are pairs of
maps between polarities, i.e., between Galois connections whose
partially ordered sets are powersets of the formal context sets.  This is
actually natural because, as stated in Section 1, the usefulness of
FCA comes from the Galois connections determined by the relations in the
formal contexts.  Thus, it is appropriate that the domains and codomains
of the morphisms in these categories of formal contexts be Galois
connections between the powersets of the formal contexts, i.e.,
the domains and codomains should be polarities.

\smallskip

The category ${\bf Gal}$
is given in \cite{MMS1987}.
In \cite{DMR2013}, a similar but different category with
formal contexts as objects is defined.  Though the work
in \cite{DMR2013} also builds on the results in
\cite{MMS1987}, the category in \cite{DMR2013} differs
from the category defined below in that both $h$ and $k$
in the definition below go from components of the
domain to components of the codomain.  In \cite{DMR2013},
the function $k$ is defined such that
$k: Q_{2} \rightarrow Q_{1}$.  This surprising direction
of $k$ actually follows naturally when generalizing
the topological systems work of S. J. Vickers \cite{Vickers}
which was done in \cite{DMR2010L} and then used in
\cite{DMR2013} to build relations between formal contexts
and systems.

\begin{defn} \rm
${\bf Gal}$ is the category whose objects are Galois connections,
$(f, (P, \leq), (Q, \sqsubseteq), g)$ and whose morphisms
\[ (h,k): (f_{1}, (P_{1}, {\leq}_{1}), (Q_{1}, {\sqsubseteq}_{1}), g_{1})
\rightarrow (f_{2}, (P_{2}, {\leq}_{2}), (Q_{2}, {\sqsubseteq}_{2}), g_{2}) \]
are such that $h: P_{1} \rightarrow P_{2}$ and $k: Q_{1} \rightarrow Q_{2}$
are functions with $k \circ f_{1} = f_{2} \circ h$ and
$h \circ g_{1} = g_{2} \circ k$.
\end{defn}

The following proposition comes from \cite{MMS1987}.

\begin{prop} \rm
If $(f_{1}, (P_{1}, {\leq}_{1}), (Q_{1}, {\sqsubseteq}_{1}), g_{1})$ and
$(f_{2}, (P_{2}, {\leq}_{2}), (Q_{2}, {\sqsubseteq}_{2}), g_{2})$ are $\bf Gal$-objects
and if $h: P_{1} \rightarrow P_{2}$ and $k: Q_{1} \rightarrow Q_{2}$ are functions, then
the following are equivalent where the conjunction of the three conditions in item 2 is
equivalent to items 1 and 3.

\begin{enumerate}

\item
$(h, k)$ is a $\bf Gal$-morphism.

\item

\begin{enumerate}

\item
$h$ maps fixed points in $P_{1}$ to fixed points in $P_{2}$,
and $k$ maps fixed points in $Q_{1}$ to fixed points in $Q_{2}$.

\item
$h$ and $k$ are level-preserving.  Thus, if $p$ and $p'$ are in the
same leaf or are equivalent in $P_{1}$, then $h(p)$ and $h(p')$
are in the same leaf in $P_{2}$, and if $q$ and $q'$ are in the same
leaf of $Q_{1}$, then $k(q)$ and $k(q')$ are in the same leaf of
$Q_{2}$.

\item
If $p \in P_{1}$ and $q \in Q_{1}$ in anti-isomorphic leaves of $(f_{1}, g_{1})$,
then $h(p)$ and $k(q)$ are also in anti-isomorphic leaves of $(f_{2}, g_{2})$.

\end{enumerate}

\item
If $x$ and $y$ are in the same leaf of $P_{1}$ (respectively, of $Q_{1}$),
then they are mapped by either path to the same fixed point of $Q_{2}$
(respectively, of $P_{2}$).

\end{enumerate}

\end{prop}


\begin{defn} \rm
${\bf Pol}$
is the full subcategory of ${\bf Gal}$
such that a ${\bf Gal}$-object is a ${\bf Pol}$-object if
and only if it is a polarity.
\end{defn}

\begin{defn} \rm
${\bf FC}$
is the category whose objects are formal contexts
and whose morphisms are those of ${\bf Pol}$.
That is, if ${\mathcal{K}}_{1}:=(G_{1},M_{1},R_{1})$ and
${\mathcal{K}}_{2}:=(G_{2},M_{2},R_{2})$ are formal
contexts, then $(h, k): {\mathcal{K}}_{1} \rightarrow {\mathcal{K}}_{2}$
is a morphism in ${\bf FC}$ if and only if
\[ (h, k) : (H_{1}, (\wp (G_{1}), \subset),  (\wp (M_{1}), \subset), K_{1})
\rightarrow (H_{2}, (\wp (G_{2}), \subset),  (\wp (M_{2}), \subset), K_{2}) \]
is a ${\bf Pol}$-morphism.
\end{defn}

\begin{prop} \rm
${\bf FC}$ and ${\bf Pol}$ are isomorphic categories.
\end{prop}

In \cite{MMS1987}, it is shown that ${\bf Gal}$ is complete, cocomplete,
well-powered, and co-well-powered  \cite{AHS}.  Thus, ${\bf Gal}$ is
a structurally rich and well behaved.
Since ${\bf Pol}$ is a full subcategory of ${\bf Gal}$ and
${\bf FC}$
is isomorphic to ${\bf Pol}$, then
${\bf FC}$ is
 seemingly ``close'' to also being structurally rich.
However, the constructions which
make ${\bf Gal}$  well behaved may not be closed with respect to
${\bf Pol}$, and thus, these constructions may not be
applicable to ${\bf FC}$.  Though this is true, we can via an embedding essentially
make the constructions closed with respect to ${\bf Pol}$
because, as shown below, each object in ${\bf Gal}$ can be embedded into
some ${\bf Pol}$-object.  Thus, we can
perform our constructions on objects in ${\bf Pol}$ by doing the construction
in ${\bf Gal}$ and then embedding the results back into ${\bf Pol}$.  Thus, in a natural
sense, we have these constructions in ${\bf FC}$, and thus, ${\bf FC}$ is structurally
rich and well behaved.

\smallskip

Given an arbitrary Galois connection, in Proposition \ref{prop-mono} we
define the ${\bf Gal}$-morphism which embeds this arbitrary
Galois connection into a polarity.  In Definition \ref{defn-embed}, we give
a precise definition of an embedding, and in Theorem \ref{thm-embed},
we prove that the morphism defined in Proposition \ref{prop-mono}
is indeed an embedding.

\begin{const} \rm
\label{const-embed}
Let $( f, (P, \leq), (Q, \sqsubseteq), g)$ be a Galois connection.
Define $F : \wp (P) \rightarrow \wp (Q)$ such that for $A \subset P$,
\[ F(A) = {\bigcap}_{p \in A} \downarrow \! f(p), \]
and likewise, define $G: \wp (Q) \rightarrow \wp (P)$ such that for $B \subset Q$,
\[ G(B) = {\bigcap}_{q \in B} \downarrow \! g(q). \]
\end{const}

\begin{prop} \rm
Let $( f, (P, \leq), (Q, \sqsubseteq), g)$ be a Galois connection.
$(F, (\wp (P), \subset), (\wp (Q), \subset), G)$ is a polarity.
\end{prop}

\noindent
{\bf Proof:} Clearly, $F$ and $G$ are order-reversing because
as the argument increases, the corresponding intersection becomes smaller.
Let $A \subset P$, and let $p \in A$.  $F(A) \subset \; \downarrow \! \!f(p)$.
Therefore, for each $q \in F(A)$, we have $q \sqsubseteq f(p)$.  It follows
for each $q \in F(A)$, that $p \leq g \circ f(p) \leq g(q)$, and thus,
$p \in \; \downarrow \! \! g(q)$ for each $q \in F(A)$.  Hence, $p \in G \circ F(A)$
for each $p \in A$, and therefore, $A \subset G \circ F(A)$.  Similarly,
$1_{\wp (Q)} \subset F \circ G$.
$\bullet$

\begin{prop} \rm
\label{prop-mono}
Let $( f, (P, \leq), (Q, \sqsubseteq), g)$ be a Galois connection.
Define $i_{P} : P \rightarrow \wp (P)$ such that
$i_{P} (p) = \; \downarrow \! p$,
for each $p \in P$.
Likewise, define $i_{Q} : Q \rightarrow \wp (Q)$ by
$i_{Q} (q) = \; \downarrow \! \! q$,
for each $q \in Q$.
\[ (i_{P}, i_{Q}) : ( f, (P, \leq), (Q, \sqsubseteq), g)
\rightarrow (F, (\wp (P), \subset), (\wp (Q), \subset), G) \]
is a $\bf Gal$-morphism.
Additionally,
$(i_{P}, i_{Q})$ is a monomorphism in ${\bf Gal}$.
\end{prop}

\noindent
{\bf Proof:}
Let $s \in P$.  It follows that $i_{P}(s) = \downarrow \! s$.
If $p \leq s$, i.e., if $p \in \downarrow \! s$, then since
$f$ is order-reversing, $\downarrow \! f(s) \subset \downarrow \! f(p)$.
Hence,
\[ F(\downarrow \! p) = \downarrow \! f(p) = i_{Q}(f(p)). \]
Hence, $F i_{P} = i_{Q} f$.  Similarly,
$i_{P} g = i_{Q} f$.  Thus, $(i_{P},i_{Q})$ is a
${\bf Gal}$-morphism.

\smallskip

From \cite{MMS1987}, we know that
$(i_{P}, i_{Q})$ is a monomorphism in ${\bf Gal}$ if and only if both
$i_{P}$ and $i_{Q}$ are injections, and since $\leq$ and $\sqsubseteq$
are both partial orders, then $i_{P}$ and $i_{Q}$ are both
injections.
$\bullet$

\smallskip

Given a Galois connection $( f, (P, \leq), (Q, \sqsubseteq), g)$,
we specify the polarity
$(F, (\wp (P), \subset), (\wp (Q), \subset), G)$ by defining
the order-reversing maps $F$ and $G$.  We could have
specified the polarity by defining the appropriate relation
from $P$ to $Q$.  From Construction \ref{const-G},
we know that this relation $R \subset P \times Q$ is
$(p, q) \in R$ if and only if
$q \in F(\{p\})$.  Thus, $(p, q) \in R$ if and only if
$q \in \; \downarrow \! \! f(p)$.  By the alternate definition of
a Galois connection, Definition \ref{defn-alt-gc},
\[ (p, q) \in R \textrm{ if and only if } p \leq g(q) \textrm{ and } q \sqsubseteq f(p). \]

\smallskip

The following four definitions come from \cite{AHS}.

\begin{defn} \rm
Let $L: {\bf X} \rightarrow {\bf Y}$ be a functor.  $L$ is
\emph{faithful} if whenever $m,n:X_{1} \rightarrow X_{2}$
are distinct morphisms in ${\bf X}$, then $L(m)$ and $L(n)$
are distinct morphisms in ${\bf Y}$.  Said differently,
$L$ is faithful if it is injective on the set of morphisms
between each two objects in ${\bf X}$.
\end{defn}

\begin{defn} \rm
Let $\bf X$ be a category.  A pair $({\bf A}, U)$ is a
\emph{concrete category over $\bf X$} if $\bf A$ is a category and if
$U : {\bf A} \rightarrow {\bf X}$ is a faithful functor.
\end{defn}

\begin{defn} \rm
Let $({\bf A}, U)$ be a concrete
category over $\bf X$.  An $\bf A$-morphism
$f: A \rightarrow B$ is an \emph{initial morphism} if
for any $\bf A$-object $C$, an $\bf X$-morphism
$g: U(C) \rightarrow U(A)$ is an $\bf A$-morphism
$g: C \rightarrow A$ whenever $f \circ g: C \rightarrow B$
is an $\bf A$-morphism.
\end{defn}

\begin{defn} \rm
\label{defn-embed}
Let $({\bf A}, U)$ be a concrete
category over $\bf X$.  If
$f: A \rightarrow B$ is an initial $\bf A$-morphism
and if $f: U(A) \rightarrow U(B)$ is a monomorphism
in $\bf X$, then $f: A \rightarrow B$ is an \emph{embedding}.
\end{defn}

In the next proposition, $\bf Set$ is the category of sets
and functions.  The only part of this proposition which we
need for our embedding result is the part involving
$({\bf Gal}, U_{G})$.

\begin{prop} \rm
$({\bf Gal}, U_{G})$, $({\bf Pol}, U_{P})$, and $({\bf FC}, U_{F})$ are concrete over
${\bf Set} \times {\bf Set}$ where
\[ U_{G}(f, (P, \leq), (Q, \sqsubseteq), g) = (P, Q), \]
\[ U_{P}(H, \wp (G), \wp (M), K) = (\wp (G), \wp (M)), \textrm{ and} \]
\[ U_{F}(G, M, R) = (\wp (G), \wp (M)). \]
\end{prop}

\begin{thm}
\label{thm-embed}
Let $(f, (P,\leq), (Q,\sqsubseteq),g)$ be a Galois connection.  The
${\bf Gal}$-morphism $(i_{P},i_{Q}):(f,g) \rightarrow (F,G)$ is an
embedding.
\end{thm}

\noindent
{\bf Proof:} \,
From Proposition \ref{prop-mono}, we know that $(i_{P},i_{Q})$
is a monomorphism.  Thus, we only need to show that it is initial.
Let $(f_{C}, (P_{C},\leq), (Q_{C},\sqsubseteq), g_{C})$ be a Galois
connection, and let $(r,s):(P_{C},Q_{C}) \rightarrow (P,Q)$ be a
morphism in ${\bf Set} \times {\bf Set}$ (i.e., $r$ and $s$ are
set functions) such that
\[ (i_{P} s, i_{Q} t): (f_{C},g_{C}) \rightarrow (F,G) \]
is a ${\bf Gal}$-morphism.  It follows that
\[ i_{Q} t f_{C} = F i_{P} s = i_{Q} f s. \]
Since $i_{Q}$ is injective, then $t f_{C} = f s$.
Similarly, $s g_{C} = g t$.  Therefore,
$(r,s)$ is a ${\bf Gal}$-morphism, and $(i_{P}, i_{Q})$
is initial.
$\bullet$




\section{Example}
\label{sect-example}

The major vision or goal driving the development of the semantic web
is to create a web which can be understood by computers
with minimal intervention by humans.  A major component of the semantic
web is RDF, resource description framework.  RDF stores information as
triples where each triple is composed of a subject, a predicate, and an object.  These
triples are simple; each triple holds only an elementary amount of
information.  However, these triples can be combined to form large graphs
representing significant and complex information.\footnote{Thanks to
S. Ayvaz and M. Aydar of Kent State University for explaining
this example.}

\smallskip

To help understand the information in a large RDF graph, it is helpful to
have a schema of the information.  The schema is a framework around
which the information may be organized.  Typically, elements of a
schema are classes of subjects, and the schema
may be organized with classes being subsets of other classes.  Not surprisingly,
a schema may be developed from information in the RDF graph itself.
One way of creating a schema is
to begin with the first two components of the RDF triples.  These
two components form a binary tuple consisting of a subject, which in
FCA terms is an object, and a predicate or property.  Each subject or
object may have many associated properties, and multiple objects may
have the same property.  Thus,
these binary subject-property
tuples form a relation between the set of subjects or objects and the set of
predicates.

\smallskip

The classes of
a schema may be, at least, partially determined
by the properties or predicates which the objects in the classes
satisfy.  Said differently, the classes of interest in forming the
schema may be related to the object set nodes determined by the Galois connection
or polarity of a formal context.

\smallskip

The semantic web is, however, not a static entity.  It is dynamic;
the RDF triples and the RDF graphs may change often.  Thus, the corresponding
schemas will, at times, need to change.  How will changes in the
RDF tuples affect the schema?  Said differently, how will changes in the
formal contexts change the schemas?  This question is a motivation for
this paper.


\section{Conclusion}
\label{sect-con}

Our work raises questions about Galois connections and
about formal concept analysis.  Though in some situations, there may be good reasons
for wanting to replace a formal context with a better formal context and though
we have shown orderings which mathematically allow us to determine ``better'' formal contexts,
this question must addressed from the perspective of FCA, i.e., from the perspective of FCA what
criteria should be used to decide when one
formal context is better than another?  Though we can mathematically define better formal
contexts, what in practice determines when one formal context is better than another?

\smallskip

When defining orderings on formal contexts, we have restricted ourselves to
sets of formal contexts which have the same set of objects and the same
set of properties; it may be useful to think about ``better'' formal contexts
when the underlying sets of objects and/or properties may change.  For example, when
working with the Semantic Web, we will likely need to enlarge our set of
objects.  Can we make these changes and preserve, at least, in part the schema structure
which is already in place?

\smallskip

In addition to understanding what makes one formal context better than
another, future work may also include understanding what the categorical
constructions in ${\bf FC}$ mean in practice.  In \cite{MMS1987}, in addition
to defining the category ${\bf Gal}$, the category ${\bf {Gal}_{p}}$ is defined.
${\bf {Gal}_{p}}$ is a subcategory of ${\bf Gal}$ such that $(h,k)$ is a
${\bf {Gal}_{p}}$-morphism if and only if it is a ${\bf Gal}$-morphism
and both $h$ and $k$ are order-preserving.  It may be worthwhile to
study the corresponding formal context category ${\bf {FC}_{p}}$.

\section{Acknowledgements}

The second author wishes to
thank Dave Schmidt for his friendship, encouragement, and research
advice and collaboration, especially, when the second
author was making a career change from mathematics
to mathematics and computer science.

\smallskip

The authors wish to thank the referees for their suggestions, including
catching some errors in an earlier draft.

\nocite{*}
\bibliographystyle{eptcs}
\bibliography{denn-melt-roda}
\end{document}